# A Simple Sinuosity-Based Method using GPS data to Support Mitigation Policies for Public Buses GHG Emissions

William Wills [a]*, Joao Meirelles [b,c]*, Vivien Green Baptista[d], Gabriel Cury[e], Pablo Cerdeira[f,c]

[a] Eos Consulting, Rio de Janeiro, Brazil
[b] Laboratory for Human-Environment Relations in Urban Systems, Swiss Federal Institute of Technology Lausanne, Switzerland
[c] PENSA - Rio de Janeiro City Hall, Brazil
[d] Energy Planning Program, COPPE, Federal University of Rio de Janeiro, Brazil
[e] GABE
[f] Center for Technology and Society, Getúlio Vargas Foundation, Rio de Janeiro, Brazil

*corresponding authors: wills@eos.eco.br, joao.meirelles@epfl.ch
*authors have contributed equally to this paper

**ABSTRACT**

It is clear by now that climate change mitigation relies on our capacity to guide urban systems towards a low-carbon phase and that the urban transportation sector plays a major role in this transition. It is estimated that around 30% of total $CO_2$ emissions worldwide come from the urban transportation sector. Regardless of its importance, detailed estimations of transport-related emissions in cities are still rare to find, hindering our capacity to understand and reduce them. This work aims to develop a replicable and fast method for GHG estimation from GPS (Global Positioning System) data and to introduce a simple sinuosity-based algorithm for such. We applied the method for 1 year of GPS data in the city of Rio de Janeiro. Our results were compared to top-down estimations from fuel consumption and proved to be valid after a simple data filling process. Our GPS-based approach allowed for much finer spatial and temporal descriptions of emissions and we further showed possible policy insights that can be extracted from the estimated emissions based on the proposed method.

**INTRODUCTION**

We live in an increasingly urban world, with more than half of the world's population living in cities and a rural-to-urban migration that is not showing signs of slowing down [1]. This urban condition profoundly changes our relation to the environment, introducing different ways of consuming, commuting and managing our basic needs. We are already facing the consequences of such transformations, with climate change quickly escalating and producing an increasing number of disasters[2]. The Intergovernmental Panel on Climate Change estimates that around 20-30% of total GHG emissions worldwide come from the urban transportation sector[3] making it one of the key topics in the low-carbon transition[4]. It is clear by now that climate change mitigation relies on our capacity to guide urban systems towards a low-carbon phase.

Regardless of its importance, detailed estimations of transport-related emissions in cities are still rare to find, hindering our capacity to understand and reduce them. City-wide inventories have been following a top-down approach, estimating emissions from the total aggregated fossil fuel consumption [17–19]. This is primarily done due to the lack of better data, but it limits the findings to coarse-grain figures, usually city-wide and year-based aggregated, providing little information for mitigation policies and improvement opportunities. This can be overcome by bottom-up approaches, using detailed activity data, such as the Global Positioning System(GPS), for determining emissions and providing spatial and temporal information not obtained by fuel consumption. Most studies analyze taxi GPS data with conclusions of little practical application by policy-makers - who need to focus on public transport data from public monitoring systems and uses high resolution monitoring systems, which require high computational and human capital.

Some recent works have been able to estimate CO2 emission from GPS data, considering that the pollution emitted by a vehicle depends on local moving parameters as well as fleet characterization and technology standards. Çagri et. al. [25] quantified several routing parameters including fuel consumption and CO2 emissions to analyze how mix-pollution routing problems can be minimized by using a heterogeneous fleet and Sun et. al. [26] proposed a trajectory-based method to estimate energy consumption and CO2 emissions in large areas using GPS data from samples of the traffic flow population. These studies involve elaborated methodologies which require fine resolution data and an elevated computational capacity, a combination of factors not easily found in urban planning agencies worldwide, especially in developing countries.

Following local idiosyncrasies, the quality of the GPS data may vary significantly. Using high-resolution GPS data, Nyhan et. al. [23] analyzed taxi trajectory from over 15,000 vehicles in Singapore to estimate CO2, NOx, VOCs and PM emissions using a microscopic emissions model in high spatiotemporal resolution, collecting GPS data every few seconds and discarding records with gaps greater than 5s. Kan et. al. [20] proposed more accurate methods for estimating fuel consumption and hot and cold start emissions by analyzing mobile and stationary activities of taxis through GPS data collected every 10 seconds on average. Luo et. al. [22] tried to understand low-carbon transportation behaviors using GPS data from taxis every few seconds, analyzing the relationship between energy consumption and CO2 emissions. However, high-resolution data is a privilege for some cities, and those working with public authorities have to make do with less.

Using low-resolution GPS data, Zhao et. al. [21] investigated CO2 emissions in taxi trajectory with spatiotemporal analysis of the emission patterns, with GPS data ranging from 10s to 60s. Shang et. al. [22] estimated fuel consumption/emissions using an adaptation of the COPERT model which considers different exhaustion modes (hot and cold) and taxi GPS data collected every 96s on average. Regardless of the resolution, concerning public policies, buses play a more important role than taxis since they are regulated by public authorities and serve a much greater part of the population. On top of that, several cities in the global south still rely on GPS data with resolutions even lower than the previously mentioned publications, such as Rio de Janeiro which had one record for every 120s on average. As far as mitigation policies are concerned, low-resolution models already provide enough insights to implement most of the mitigation policies, making it important to find an easier yet reliable way of assessing GHG emissions of the public transport sector through a bottom-up approach in a more straightforward method.

This work aims to develop a simple method for GHG estimation from low-resolution GPS data using vehicle kilometers traveled (VKT), a bottom-up approach with no demand for complementary data or complex methodology. The model aimed at being replicable and straightforward so it can become a tool in decision making and planning in the public transport sector, identifying and differentiating local and general problems and opportunities with simple assumptions and methodology.

The public bus fleet of the city of Rio de Janeiro was used as the case study with no other dataset but the GPS data and the street network. The results have been compared with the official top-down estimates, and they are validated after a simple data-filling process. Afterwards, we showed possible policy insights that can be extracted from the estimated emissions using spatial-temporal information yielded by the proposed method.

## DATA AND METHOD

### Database

We collected the Global Positioning System (GPS) records for every public bus for the city of Rio de Janeiro between Jul-2014 and Mar-2016 from the municipal open data portal (data.rio). Information on the geographical position (lat, long) of the buses, date-time, route number, unique ID and instantaneous velocity of the buses can be retrieved for each GPS record from the portal. The database comprises on average one record every 2 minutes for all the working buses, which makes a daily occurrence of between 3-5 million records. This time interval can be defined as low-resolution for GPS data [20] and, as mentioned before, it is significantly longer than the time-interval found in other publications, even among low-resolution GPS studies [21,22]. Furthermore, a lot of gaps can be observed in the transmission of GPS and some days would experience very few records.

Data was streamlined with a python scraper code to a PostgreSQL columnar database into a data server. The database had more than 2 billion records during the 20 months of collection, heterogeneously distributed across the days. Because of the unequal distribution and to make it comparable to top-down calculations, a period of one year was selected from the database to be used: Dec-2014 to Dec-2015. The algorithm developed to produce the analysis was implemented in PostgreSQL + PostGIS and consisted of the steps described below.

### Methodology

The most straightforward method to calculate GHG emissions from GPS would be to figure, for each pair of records, the traveled distance and, using the time spent to travel the computed distance, estimate the velocity and use it as a variable in the functions for consumed fuel and GHG emissions. However, the size of our database makes this methodology computationally impossible. The traveled distance, calculated using shortest path algorithms have high computational costs, and it was necessary to find a way to perform the calculation within an acceptable time-window if we wanted to make the method replicable. To solve this problem and reach a fast and replicable method, we decided to calculate an average sinuosity between two sequential GPS records for a representative sample and then we used this value as a corrector factor to the rest of the database to estimate the driving distance of buses. From the corrected driving distance and the time spent between a sequential pair of GPS, we estimated fuel consumption and GHG emissions for that specific pair of GPS. Finally, we used a method to fill missing data.

The GPS database had some data gaps in the form of days or hours with the reduced record. The failures were due to the numerous periods when the municipal system of collection-emissions was entirely off, or when a connection problem in the recording server impeded the data to be received or recorded correctly. Thus, in order to validate our method, it was necessary to develop an algorithm to fill in the missing information.

We filled the days in which there were failures in the transmission, reception or recording of the information in the Database. For each weekday (Mon, Tue, Wed..) we defined an expected range of pairs of GPS as the bottom and top limits of the tenth decile in the distribution. For days with fewer pairs of GPS than its expected range, we filled in synthetic pairs of GPS. The results were validated with top-down GHG calculations, using fuel consumption by the municipal bus fleet.

The sinuosity-based methodology developed is composed by the following steps (Figure 1): i) For a statistically significant sample of records, an average sinuosity value was calculated, being the sinuosity the division of the reconstructed distance between two sequential GPS records and the linear distance between them (Fig 2); ii) For every sequential record in the database with a time interval below a limit value ($\Delta T$), the linear distance between all sequential records was calculated, which was corrected using the sinuosity value found in step i; iii) for each sequential pair of GPS, the average speed of the bus was calculated, thus avoiding the use of the instantaneous velocities measured by GPS devices; iv) with the calculated speed and distance data, a speed consumption function was used to infer the fuel consumption and; v) from this consumption the total emissions were calculated. Every step is further described below:

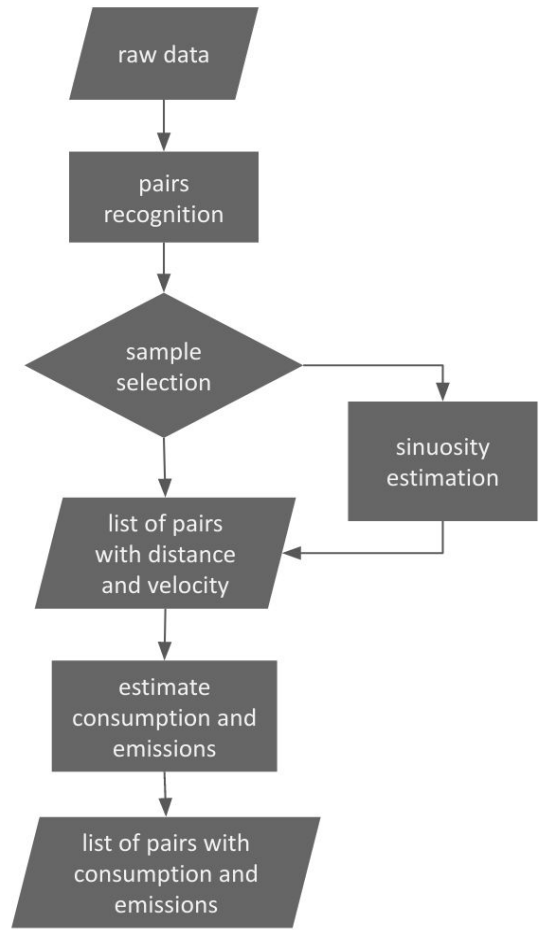

Figure 1. Sinuosity-based method algorithm scheme

**GPS pairs recognition**

After the first cleaning of absurd GPS records (outside of geographical boundaries and above 120 km/h), sequential pairs of GPS records were recognized as the GPS pairs that are emitted from the same vehicle within a time interval of fewer than 3 minutes. The interval of time is necessary to ensure a good approximation of reality. If this parameter were relaxed, we could infer distances very different from the real distances.

**Sinuosity**

For an effective calculation of CO2e emissions, it was necessary to estimate the distance traveled by each bus in the time interval studied. Determining a distance traveled from GPS records with low-resolution is not a trivial task since the path reconstruction algorithms (shortest-path) are very computationally costly. In this way, we chose to select a representative random sample (1%) of the universe of GPS pairs and reconstruct their way using shortest path algorithms (Figure 3). The shortest path algorithm was used due to the lack of an official bus route database (GTFS), which would have been the best option for this task. The proximity between pairs of GPS ($\Delta T<3min$) helps to ensure that the shortest path assumption is close to the official route. Thus, the real distance traveled (RD) of the bus within the time interval of the pair of records was calculated. The sinuosity value was computed as the simple ratio between the Real

Distance and the Euclidean Distance (RD / ED). Figure 2 presents a schematic description of this process.

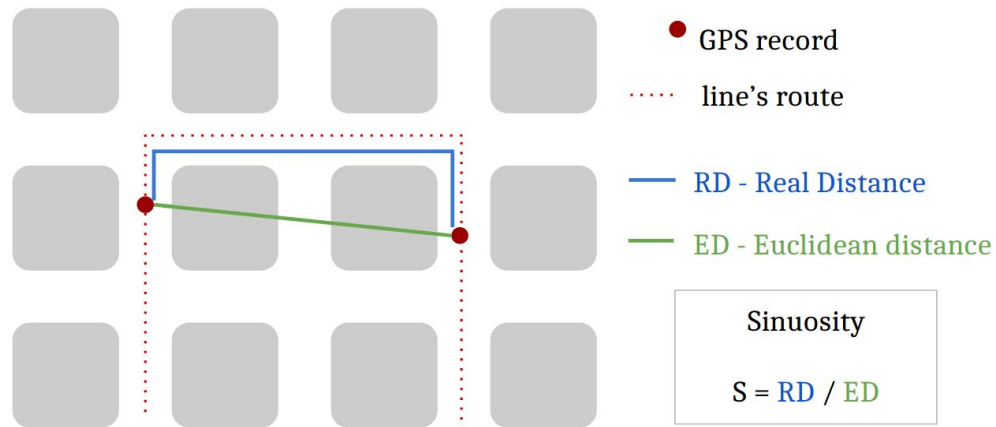

Figure 2. Methodology description. The sinuosity, a corrector factor, estimated from a representative sample of the database is a simple ratio between the Real Distance and the Euclidean Distance of two sequential GPS records. Real distance was estimated from the shortest path algorithm.

The shortest-path algorithm produced a residual number of absurd paths. Something like 0.1% of the GPS pairs in the sample resulted in extensively traveled paths for minimal time intervals. For certain pairs, the estimated speed of the bus was much larger than would be physically possible for the vehicle. These paths were removed from the base using a speed filter (> 120 km/h). Other paths resulted in distance traveled value equal to zero. This happened because of the way the algorithm works, always starting and ending in a street corner. Thus, in the cases where the two sequential GPS records were emitted very close (<50 meters), the origin of the two points of the pair was the same, and the sinuosity was therefore zero. In these cases, the linear distance was adopted as the final distance. A histogram of sinuosity values found for our sample of pairs of GPS can be found in Figure 3.

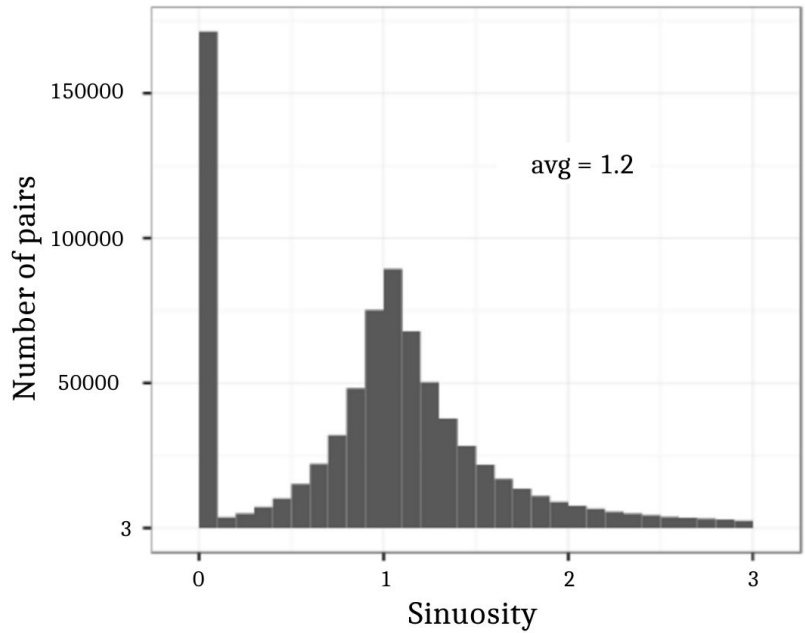

Figure 3. Probability distribution of calculated sinuosity

For estimating the real distance in the full database, the average sinuosity value S found for our sample (1.2) was used as a multiplier of the linear distance between each sequential pair of GPS. For those pairs with linear distances smaller than 50 meters, the real distance was assumed to be equal to linear distances. Thus, one arrives at a database of pairs of sequential records, each containing a distance traveled and a time interval spent.

### GHG Emissions Estimation

For each pair of GPS, the fuel consumption was calculated from an average speed-dependent estimation model[21]. The model takes into account the distance traveled and the bus speed during the time interval to estimate fuel consumption from a speed-dependent function. Although multiple factors affect fuel consumption in vehicles [20], we adopted a simpler model because no further information was available, such as bus type or fabrication year. The adopted consumption function was obtained from the European Environment Agency[21]. It is worth noticing that the curve relating consumption to the speed range of the vehicle was drawn from international research, prepared for a relevant generic technology at the time it was performed and in conditions different from those presented in the city of Rio de Janeiro. The elaboration of a consumption curve by a customized speed range for the studied city by circulating bus would also be of great importance and would allow even more precise estimations of consumption and GHG emissions. Finally, in order to estimate GHG from the consumed fuel, we adopted national emissions factors [22]. We assumed emissions factors for Diesel B6 (2.51 tCO2e/m3) or B7 (2.49 tCO2e/m3), depending on the analyzed period. Thus, the CO2e emissions of each sequential pair of GPS were calculated for each bus, each municipal line, exhaustively.

## RESULTS AND DISCUSSIONS

### Validation

In order to validate our method, we compared our results to a top-down GHG emissions estimation based on fuel consumption, the one currently used in official estimations for the city[14]. Data on diesel consumption, kilometers traveled, and passengers transported by buses in the city were obtained from the Federation of Passenger Transportation Companies of the State of Rio de Janeiro (FETRANSPOR). A later adjustment was performed in order to fill in the missing data, as detailed in the methodology. The results of these adjustments are presented in the following graphs (Figure 4), which compares Total Mileage, Diesel Consumption, and GHG Emissions per month analyzed with the top-down estimation.

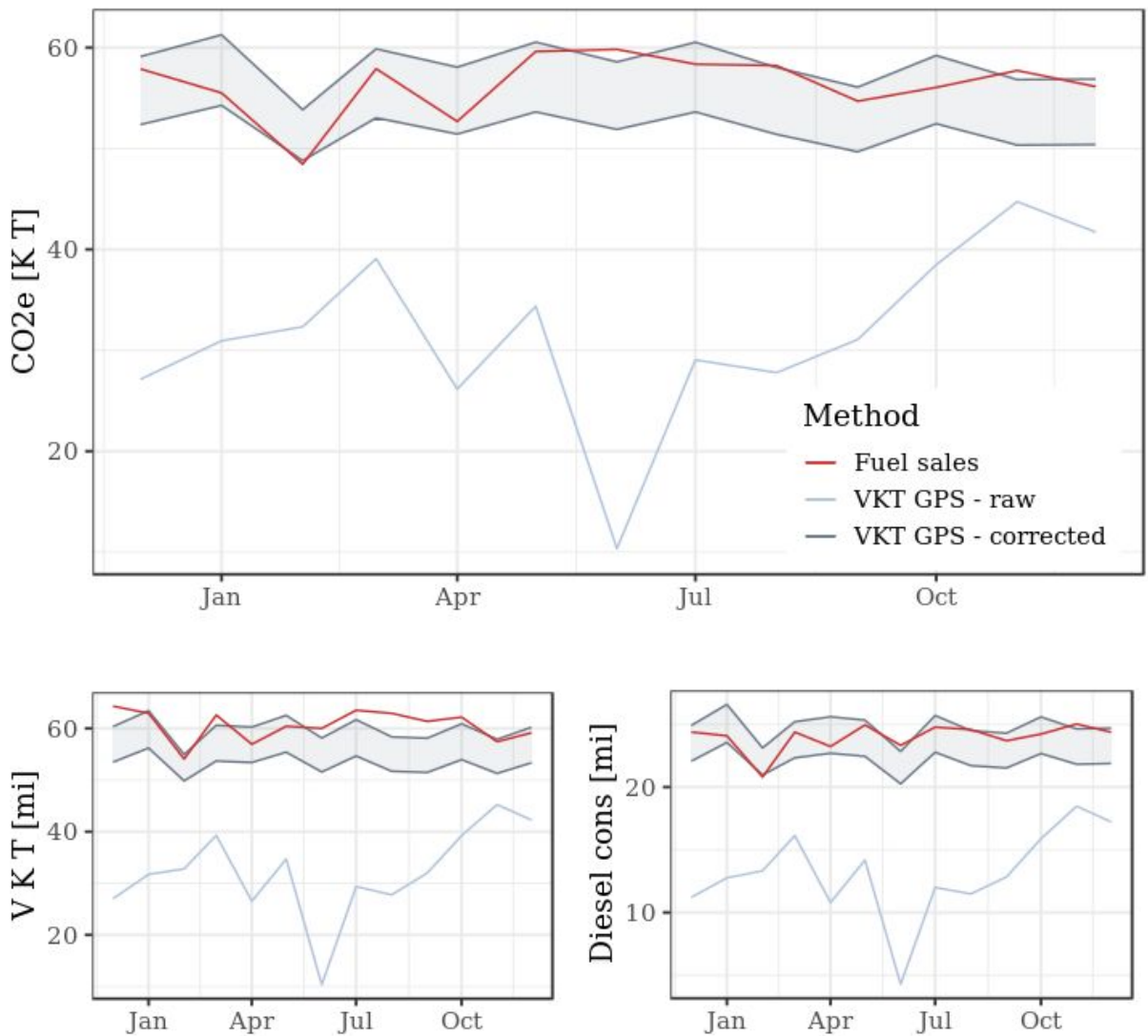

Figure 4. Validation against "top-down" methodology (red lines) for three different measurements – Km traveled, diesel consumption and CO2 emissions. The light blue line presents the raw GPS-based calculation and the grey band is the corrected confidence interval for the GPS-based calculation. Data ranges from Dec-2014 to Dec-2015.

The results obtained directly from the proposed method, without any correction, present a significant difference in all the analyzed months, and especially at the beginning of the series, which from October 2015 started counting on a more substantial number of points collected per day. By filling in the missing information on the Database as described above, we see that the order of magnitude of the numbers becomes equivalent and that our estimated range agreed with top-down estimations, with only a few months out of the predicted range, and with insignificant divergences. The method is further validated by it's robustness to better data sample: by the end of the studied period, more records were collected and fewer days needed to be filled in. Even so, the estimated values continued to be very close to the official, top-down estimation.

Part of the observed differences can be easily explained by the use of a generic speed-consumption curve drawn from an international study. With a speed-consumption curve explicitly developed for the buses of the city of Rio de Janeiro.

### Policy implication

Our method generated fine-grained spatial-temporal profiles of emissions (Figure 5). On the left of Figure 5, an emissions map shows the spatial distribution of total greenhouse gas emissions on a Tuesday, November 24, 2015, when a large number of data points were available. Emissions were aggregated on a square lattice over the city map. This example is a good sample of the behavior of GHG emissions on a typical business day in the city of Rio de Janeiro. This type of map can be generated for different periods and sets of days, according to the purpose of the study to be done. The routes with higher GHG emissions were necessarily places where there were higher fuel consumption and emissions of local pollutants such as particulate matter and carbon monoxide. In this way, the proposed methodology can help in the orientation of public policies that aim to improve urban mobility and whose immediate benefits are the reduction of traffic, average travel time in the city, fuel consumption and, consequently, emissions of GHG and local pollutants.

The accumulation of emissions can be seen on some main roads, such as Avenida Brasil, Avenida Presidente Vargas, Linha Amarela and Avenida das Américas, among others. Of these, Avenida Brasil and Avenida das Américas are receiving BRT corridors. In a more dispersed way, it is possible to verify a lot of roads in the North Zone with emissions that are not negligible. This result may point to the need for investment in more efficient mass transportation in the region, such as a BRT corridor, a VLT or a Subway in the East-West direction, to complement the existing underground and train lines. The temporal profiles (Figure 5 below) are as well as essential decision-making tools. They can indicate optimum hours for temporal dedicated bus corridors or reduced public transportation fares.

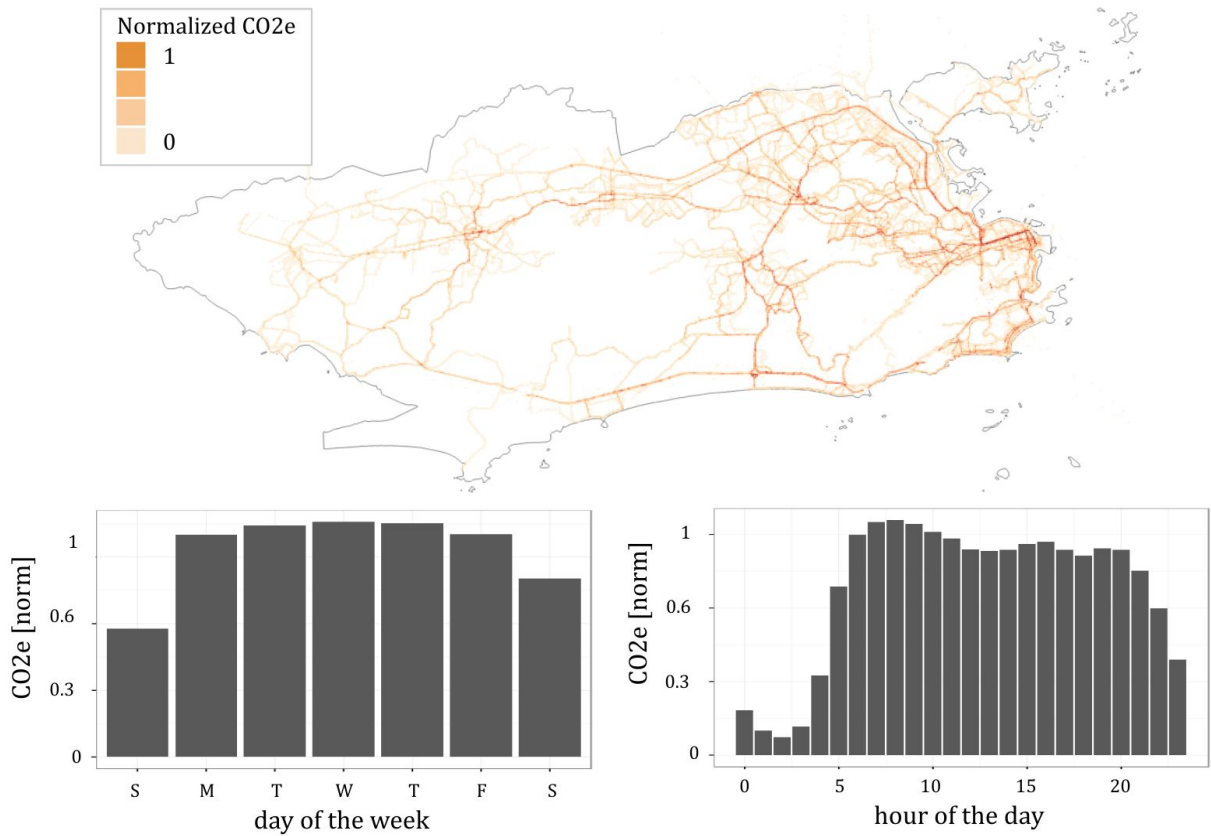

Figure 5. Spatial-temporal distributions of predicted daily emissions from the buses' GPS data. The map in the left indicates average normalized emissions - CO2e / max(CO2e) - per road link for Tuesdays. Graphs on the right show the average temporal distributions of emissions per day of the week and per hour of the day. Only days within the first decile were considered.

The method also made possible to recognize the most emitting bus lines. Figure 6 shows the distribution of GHG emissions by lines (Fig 6 above), and it can be seen that a small number of lines emit much more than the rest. These lines would be the primary candidates to undergo interventions, such as measures to alleviate traffic on their route, or even the creation of a dedicated bus corridor, or, depending on the daily flow of passengers, a segregated bus corridor or subway. Figure 6 also shows the spatial distribution of these most emitting lines (Figure 6 below). It is possible to observe that those lines are very much concentrated on specific roads/regions of the city. These roads/regions could be considered as priorities in the formulation of public policies, with measures aimed at reducing traffic, and consequently the travel time, diesel consumption, and emissions of local pollutants and GHG.

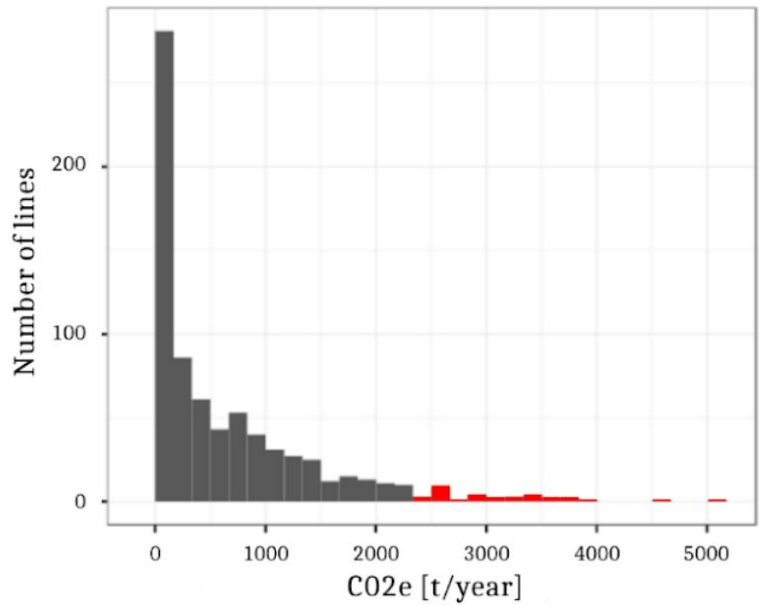

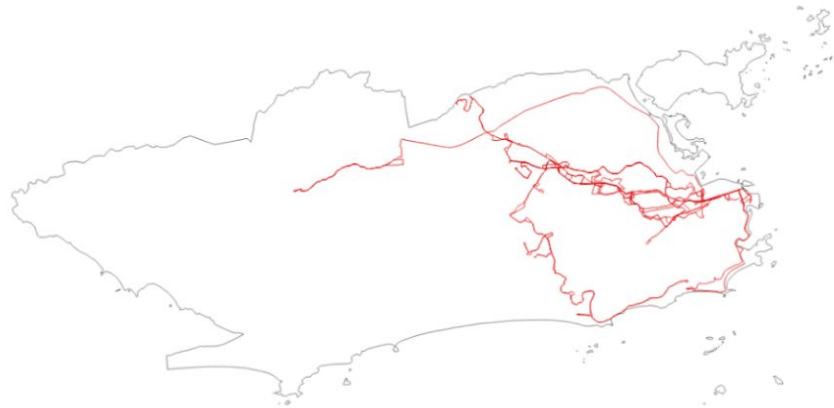

Figure 6. Distribution of CO2e emissions by line (above) and a map of most emitting lines (below). Few lines are responsible for most emissions of the public buses in Rio de Janeiro. Those lines are localized in the same region of the city.

As a next step, we tried to understand the impact of traffic on the CO2 emission of buses via a FreeFlow analysis. The Freeflow analysis aims to estimate the real effect of traffic on average speed, fuel consumption and GHG emissions from bus lines in the city of Rio de Janeiro. This analysis consists of comparing, for the same line, the difference between these parameters when the bus is running at dawn (from 00h to 03h) and when it is running at peak hours (from 08h to 12h).

Figure 7 shows a map of lines whose emissions are most impacted by traffic (10% more affected in red) and lines whose emissions are less impacted by traffic (10% less impacted in green). These results can help to guide public policies - the creation of dedicated bus lanes in the roads where the most impacted lines travel would have a significant impact on reducing diesel consumption and on GHG emissions and local pollutants.

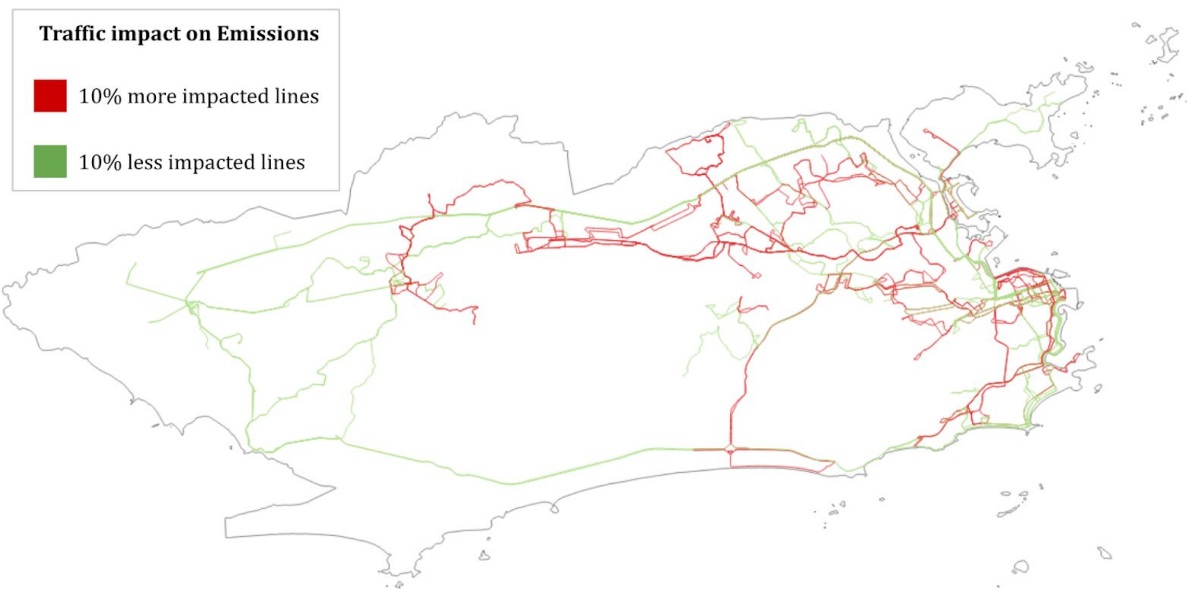

Figure 7: Freeflow Analysis - Lines most impacted by traffic (in red) and lines less impacted by traffic (in green) on a typical Tuesday.

## CONCLUSIONS

In this study, we proposed a simple and fast method for GHG emissions estimation from GPS data of buses based on sinuosity. We validated our findings against a state-of-the-art top-down method based on fuel consumption. After a simple data filling algorithm our results were found to be very similar to the top-down approach, but with a much finer spatiotemporal description, allowing for better public policy orientation.

The emissions maps present highlighted roads where diesel consumption and greenhouse gas emissions from municipal buses are concentrated. These points generally coincide with the high emission points of local pollutants, such as particulate material and NOx. This map, plotted for a typical Tuesday, can be plotted for cumulative emissions per month, or per year, and can serve as input for public policymaking aimed at reducing the emission pollutant gases locally and globally.

From the analysis of the results, it was possible to verify that the emission distribution across bus lines follows a Pareto-like shape, with some specific lines emitting a major percentage of total emission in the city. These same lines, when traced in the city map, also present the most problematic routes regarding traffic intensity, and consequently, fuel consumption and GHG emissions, indicating a clear target for mitigation policies such as electrification, express bus corridors, or, depending on the flow of passengers, a dedicated bus corridor, tramway or subway.

The methodology presented here was first developed within Rio de Janeiro's city-hall, by policymakers. Thus, it was designed after demand and had direct implications on the city's planning processes. We aim at making the simple yet useful methodology available for other practitioners worldwide, but it should be taken as a solution that emerged locally and should be tested and validated before adopted by other cities.

In order to get even better results, a database with less disruption would be important. On top of that, our method would benefit if the GPS data also contained more detailed information on the size and production year of the bus, as well as the number of passengers on board at any given moment (weight transported). These parameters would allow a much more accurate estimate of fuel consumption. The information on the number of passengers at each moment, besides contributing to the calculation of consumption, would open a new set of possibilities for studies and analyses. It would be possible, for example, to know the most and least efficient lines regarding consumption and emissions per passenger, as well as to verify how these indicators would behave by the hour of the day, by day of the week and by the month of the year, in each one of the lines.

In this research, the curve that relates the consumption to the speed range of the vehicle was drawn from an international study, prepared for a relevant generic technology at the time it was performed and in conditions different from those presented in the city of Rio de Janeiro. The elaboration of a consumption curve per speed band customized for the city of Rio de Janeiro, by circulating bus, would also be of great importance and would allow an even more accurate calculation of consumption and GHG emissions.

It is also possible to calculate emissions of various local pollutants from emissions curves of these pollutants by speed range. If these curves are available for the city of Rio de Janeiro, this would be another outstanding contribution of the tool in the orientation of public policies.

**Acknowledgments**

This study is a result of the Low Carbon Program of the city of Rio de Janeiro, an initiative of the World Bank in partnership with the City Hall and financed by the Korea Green Growth Trust Fund (KGGTF). The authors would like to thank the city hall of Rio de Janeiro and FETRANSPOR for supporting this publication with data and infrastructure.